\documentclass{ws-mplam}
\usepackage[super,compress]{cite}
\usepackage{graphicx}
\usepackage{hyperref} 
\usepackage{CJKutf8} 
\usepackage[utf8]{inputenc}
\usepackage[T1]{fontenc}
\def\draftnote{}
\begin{document}
\markboth{Y. Shi}{Wu and quantum entanglement}
%
\catchline{40}{9\&10}{2025}{2530001}{6} 
%

\title{Chien-Shiung Wu as the experimental pioneer
\\ in quantum entanglement: a 2022 note
}

\author{Yu Shi}

\address{Wilczek Quantum Center, Shanghai Institute for Advanced Studies, Shanghai 201315, China \\Department of Physics, Fudan University, Shanghai 200433, China\\
yushi@fudan.edu.cn }

\maketitle

\begin{history}
\received{11 February 2025}
\accepted{12 February 2025}
\published{21 March 2025}
\end{history}

\begin{abstract}
Advised by Prof. Chen Ning Yang, we review the early pioneering work by Chien-Shiung Wu on entangled photons created from the electron-positron annihilation. 
This paper  formed a  basis of   the author's speech at International Symposium Commemorating the 110th Birth Anniversary of Chien-Shiung Wu, May 31, 2022, and of a few  articles on the subject  in Chinese.
\end{abstract}


\ccode{PACS numbers: 03.67.Mn, 12.20.Fv, 01.65.+g}



\section{Early research experiences of C. S. Wu }

From 1930 to 1934, Chien-Shiung Wu received her undergraduate education in Central University (renamed as Nanjing University in 1949), located at Nanjing, then capital of China. Her thesis was on testing the Bragg equation of x ray diffraction in a crystal, under the supervision of Shih-Yuan Sze (1908-2007), who in 1933  became a professor and the head of the physics department at this university,  having obtained his D. Sci under the supervision of Madam Curie in 1933, working on nuclear spectroscopy.

After graduation, Wu worked in Zhejiang University at Hangzhou for a year as a teaching assistant, before joining Institute for Physics, Academia Sinica, then at Shanghai, where she worked on spectroscopy of gases with Z. W. Ku, who had received her PhD under the supervision of   D. M. Dennison at University of Michigan, supported by Levi Barbour Scholarship. In a year, likely with the recommendation of Ku, Wu prepared to go to University of Michigan for graduate study with self support provided by her uncle. En route, in visiting Physics Department of the University of California at Berkely, she was impressed by this department, especially the cyclotron, invented by   Ernest Lawrence.  Besides,  she also heard of some disadvantages  in the University of Michigan, including the discrimination against women as well as the prospect of socializing mainly with fellow Chinese students as there are hundreds of Chinese students there,   she decided to remain in Berkeley for her graduate study, and she succeeded \cite{wubio}.

From 1936 to 1940, Chien-Shiung Wu  did her PhD in Berkeley~\cite{wubio}. Her research started in 1938, with   Ernest Lawrence   and Emilio Segr\`{e} as  her advisors.  Wu was the first student of Segr\`{e} in USA.

Her first experiment, in 1938, supervised by Lawrence,  was on the x rays excited by electrons from beta decay of the radioactive lead.

Starting in 1939, supervised by Segr\`{e} ,  Wu worked  on the products of uranium fission, especially xenon. From 1940 to 1942, Wu remained here  as a postdoc, continuing the work on the fission products  of uranium and other elements, often using neutrons produced in the cyclotron.

In 1942, Wu left Berkeley to teach in  Smith College, Massachusetts. In 1943, she became a lecturer in Princeton University. In 1944, she joined Columbia University as a  member of the Manhattan Project. She joined the processing of concentrated uranium, and her main focus was on developing very sensitive gamma ray detector~\cite{wubio}.

Her result in Berkeley  on the  cross section of the absorption of neutrons by xeron became  useful in the Manhattan Project.

After the War, Wu remained in Columbia first as a research scientist. Starting from 1946, she published a few papers on slow neutron scattering of hydrogen. Starting from 1949, she worked on beta decay  and became an expert on it.

\section{Wu-Shaknov Experiment}

In 1949, Wu and her student Irving Shaknov studied the angular correlation of scattered photons from electron-positron annihilation~\cite{wu-shaknov}.

The Wu-Shaknov experiment confirmed an important theoretically  predicted value. The experiment originated in a proposal by   John Wheeler  in 1946,  to
verify the prediction of quantum electrodynamics that the  two photons emitted in the annihilation of a positron and an electron, with zero total angular momentum, must be polarized at right angles to each other. In Wheeler's proposed experiment, each photon is scattered by an electron, and  coincidence measurements are made on the two photons scattered at a same scattered angle, which is polar angle with respect to the initial direction of the photon,  but with different azimuthal angle differences.   A key quantity is  the asymmetry ratio  between the coincidence counting rates for parallel scattering directions and  perpendicular scattering directions.

The detailed theoretical investigations,  made by Pryce and Ward, and  by  Snyder, Pasternack, and Hornbostel,  predicted maximal asymmetry ratio   to be  2.85, at a scattering angle of 82 degree. Then using two Geiger  counters as detectors,  Bleuler and Bradt had observed an asymmetry ratio not inconsistent with the theory, but with the error so large that it cannot be compared with the theory in details. Similar experiments by Hanna using more efficient counter arrangements gave the  asymmetry ratio consistently smaller than predicted.

Thus an experiment based on
more efficient photon detectors and more favorable conditions were needed. This Wu and Shaknov provided. Their paper was published
as a Letter to the Editor  in 1950.

In their paper, Wu and Shaknov wrote: ``the recently developed scintillation counter has been proved  to be a reliable and highly efficient gamma-ray detector.'' They improved the   efficiency to be around ten times that of Geiger counters, consequently the coincidence counting rate increased 100 times. They used two photomultiplier tubes and two  anthracene crystals, where the photons are scattered. The positron source Cu$^{64}$  was activated by deuteron bombardment on a copper target in the Columbia cyclotron. In their experiment, the mean scattering angle is very close to 82 degree, the predicted maximum of anisotropy. In taking the coincidence measurements, one detector was kept fixed, and the other was oriented to  be with azimuth differences  of 0, 90, 180, and 270 degrees.  The asymmetry ratio  was found to be $2.04 \pm 0.08$, which was very close to the theoretical value  $2.00$.

\section{Quantum entanglement}

The Wu-Shaknov experiment was the first one producing quantum entanglement in a controlled manner. Note that we have specified ``in a controlled manner'', as most of the quantum states in nature are entangled. But entanglement produced in a controlled manner allows clear  investigations on entanglement.

Investigation on entanglement had been pioneered by  Einstein, Podolsky and Rosen (EPR) in 1935~\cite{EPR}. In their thought experiment on two particles entangled in positions  or momenta,  EPR exposed that in such a quantum state, unless there are elements of reality beyond quantum mechanical description,  there must exist correlation between particles, however far from each other. EPR argued that  this means that quantum mechanics is not a complete description of reality. The quantum  correlation exposed by EPR  was dubbed  quantum entanglement by Schr\"{o}edinger in the same year.

In his response to  EPR paper, Niels Bohr insisted that the entangled particles are inseparable, however far apart, and are also indivisible with the measurement apparatus interacting with one of the particles~\cite{bohr}.

In the following year, Wendell Furry proposed that after the separation of the entangled particles,  in the sense that their wave functions do not overlap, the entangled state reduces to a product state in which each particle is in a definite state~\cite{furry}. One may further suppose a classical probabilistic mixture of various kinds of product states. Clearly, for a uniform probabilistic mixture,  a measurement  of each individual particle  gives the same result as the entangled state. On the other hand, argued Furry, such a classical probabilistic mixture seems to  avoid the problems raised by EPR.  In a private communication with David Bohm, Einstein himself proposed that many-body quantum mechanics may break down when particles are far enough part.

In 1951, Bohm gave a technically simpler version of EPR argument, using an entangled spin singlet state of two spin-half particles~\cite{bohm}. Bohm maintained that the combined system is conceptually analyzable into components that satisfy appropriate laws. He  developed a hidden variable theory, which is causal, realistic but with hidden  nonlocal interaction between distant particles, and admittedly  artificial in form. 

\section{Bohm-Aharonov revisiting The  Wu-Shaknov experiment}

In 1957, Bohm and his student Aharonov noted that at that time,  the kind of  entanglement as in a  spin  state can only be experimentally studied in terms of entangled states of  photon   polarizations, as produced in positron-electron annihilation~\cite{bhom-aharonov}. Bohm and Aharonov still used the word ``correlation'' rather than ``entanglement''.  Previously people had already noted the actual  quantum state of the photon pair, though without attention to its significance of entanglement~\cite{heitler,snyder}.

Bohm and Aharonov made a thorough investigation on the crucial role of  quantum entanglement in the photon pair in producing the result of  the coincidence measurement after each photon is scattered by an electron, i.e. in a Compton scattering. They demonstrated that only the entangled state produce the correct theoretical value very close to the experimental result of  Wu and Shaknov, while the products or their classical probabilistic mixture, as proposed by Furry, lead to very different calculation results.

Therefore, thanks to the analysis of Bohm and Aharonov,  The Wu-Shaknov experiment done a few years earlier indeed produced an entangled state of photon polarizations, and became a smoking gun showing that EPR discussions concern real properties of matter, rather than ``mere philosophy''.

\section{An attempt in testing violation of Bell Inequality}

In 1964, John Bell published an inequality that later bears his name, which is a consequence of the assumption of locality and realism but is violated by quantum mechanics. Bell inequality concerns correlations between the entangled particles. In order that the Bell inequality is violated, the two spins need to be measured along two directions that are neither parallel nor perpendicular.

It was considered whether Wu-Shaknov experiment, or some variation of it, can be used to test Bell inequality. However, there are two obstacles.  One, as noted by Abnor Shimony and John Clauser respectively, was that in Wu-Shaknov experiment, polarizations in the two wings of the apparatus are either parallel or perpendicular, and could not be  in other angles. Clauser visited Wu, who confirmed this.   The other problem was that even if the first problem were put aside,   the photons were measured by scattering with electrons, but transfer from the entanglement between photons to electrons is too weak to allow violation of Bell inequality. On the other hand, the  energies of the  photons produced in electron-positron annihilation are too high to be measurable by using polaroids,  thus they would go through without being measured. Such a measurement is only suitable for low-energy photons, as nowadays routinely done in laboratories.

Clauser's visit raised Wu's interest in this subject. With her student Leonard  Kasday and John Ullman, she tried an experiment testing the violation of Bell inequality~\cite{kasday}.

They analyzed their result by making two additional assumptions. First, it was assumed that an ideal polarization analyzer can be constructed. Second, the results in the ideal polarization analyzers and in Compton scattering are described by quantum mechanics anyway. Then they argued that their experiment agreed with quantum mechanics and violated local realism. It was generally agreed that these assumptions are not serious.

\section{Summary}

The Wu-Shaknov experiment is historically very important, as it establishes the validity of quantum entanglement and making Wu the experimental pioneer in it, even though at the time of experiment these were not known.

The success of this experiment must have been benefited from Wu's   experience in developing very sensitive gamma ray detector for Manhattan project.

Wu's work was characterized by precision. The most famous testimony of it was of course the Wu experiment on parity. Nevertheless,  the Wu-Shaknov experiment was also a testimony.

\section*{Acknowledgment}

Quite a few years ago, Professor Chen Ning Yang  advised me to include Wu's work on quantum entanglement when writing about Chien-Shiung Wu. The author thanks Prof. Yang  for   his advice, encouragement and discussions through the years.   

This paper, except the abstract and acknowledgment, was written in 2022. It was sent to Prof. Da Hsuan Feng on May 16, 2022.
2022, and I mentioned that I would publish it in Modern Physics Letters A, Feng said nobody would read it, I replied that it could also be published in Physics Today.  The paper formed the basis of my speech ``Scientific Spirit of Chien-Shiung Wu: From Quantum Entanglement to Parity Nonconservation'' made in International Symposium Commemorating the 110th Birth Anniversary of Chien-Shiung Wu, May 31, 2022~\cite{abstracts,videos}, which  was followed by a few Chinese articles~\cite{chinese1,chinese2,chinese3,chinese4}.

\end{document}